# Revaluation of the lower critical field in superconducting $H_3S$ and $LaH_{10}$ (Nature Comm. 13, 3194, 2022).

V. S. Minkov[1], E. F. Talantsev[2,3], V. Ksenofontov[1], S. L. Bud'ko[4,5], F. F. Balakirev[6], M. I. Eremets[1]

[1]*Max Planck Institute for Chemistry; Hahn Meitner Weg 1, Mainz 55128, Germany*
[2]*M.N. Mikheev Institute of Metal Physics, Ural Branch of the Russian Academy of Sciences, S. Kovalevskoy St 18, 620108 Ekaterinburg, Russian Federation*
[3]*NANOTECH Centre, Ural Federal University, Mira St 19, 620002 Ekaterinburg, Russian Federation*
[4]*Ames National Laboratory, U.S. Department of Energy, Iowa State University; Ames, IA 50011, United States*
[5]*Department of Physics and Astronomy, Iowa State University; Ames, IA 50011, United States*
[6]*Los Alamos National Laboratory; Los Alamos, NM 87545, USA*
*Corresponding authors:
Email: m.eremets@mpic.de (M.I.E.); v.minkov@mpic.de (V.S.M)

In our paper[1], we studied the magnetic response of $H_3S$ and $LaH_{10}$ superconductors to an applied magnetic field using Superconducting Quantum Interference Device (SQUID) magnetometry. Hirsch, in his comment[2], highlighted an inconsistency in the data averaging procedure while questioning whether high-$T_c$ hydrides are superconductors at all. We accept the criticism regarding our method of extracting the penetration field $H_P$ from the original data. Our SQUID magnet becomes noisy at high magnetic fields, which necessitated the smoothing of a small portion of the data. To eliminate any data processing issues, we have performed an alternative data analysis that does not require data smoothing to estimate the penetration field $H_p$ values. The formulation of the analysis is identical to the one widely used for determining critical currents in superconductors[3]. Recently, it has been shown to work effectively for extracting $H_p$ and the lower critical field $H_{c1}$ from *DC* magnetization data[4]. The $H_p$ values of the present analysis are consistent with those published in our original work[1]. We wish to emphasise very clearly that the criticism pertains to the secondary issue of determining Ginzburg-Landau parameters for these hydride superconductors and does not undermine the validity of the existence of hydride superconductivity. Indeed, as part of our paper[1], we also published *m(H)* curves demonstrating the virgin curve (about which the analysis issues were raised) followed by magnetic hysteretic loops that have the classic form of the hysteresis curves of superconductors. Above $T_c$, in both $H_3S$ and $LaH_{10}$, the hysteresis is absent. We make all the data available so that readers can judge for themselves.

First, we take the opportunity to briefly review the status of work on hydride superconductivity, responding to inferences by Hirsch and co-workers that it is somehow not real.

The discovery of high-temperature superconductivity in hydrogen sulfide at $P \approx 150$ GPa, with a record critical temperature ($T_c$) of 203 K[5], marked a milestone in superconductivity research and established a new family of materials known as hydrogen-rich superconductors[6,7]. This family includes hydrides with different types of chemical bonding – covalent and metallic – between hydrogen and heavy atoms. All exhibit conventional superconductivity, allowing accurate $T_c$ calculations for certain crystal lattices. Advances in crystal structure prediction have significantly accelerated the search for new, promising high-$T_c$ superconductors[7].

Despite the challenges of probing and studying superconductivity in samples with a size of tens micrometres within diamond anvil cells under extreme pressures, the existence of superconductivity in hydrogen-rich compounds is firmly established by various independent experimental techniques and independent research teams[8]. These compounds have achieved $T_c$ values reaching 245 K in $YH_9$[9,10] and 250 K in $LaH_{10}$[11-14].

Hydrogen-rich superconductors exhibit all necessary hallmarks of superconductivity, which have been probed under megabar pressures, and different research groups have well reproduced their properties[8]. High-temperature superconductivity at high pressures is evidenced by:

1. Zero electrical resistivity below $T_c$, or at least orders of magnitude lower than that of any known metal;

2. The Meissner state, involving the expulsion of an applied magnetic field from the sample upon cooling below $T_c$;
3. The characteristic complex response of a superconductor to applied magnetic fields, which differs between type-I and type-II superconductors. For type-II superconductors, several phenomena are associated with the penetration of magnetic fields into the sample and include trapped magnetic flux, vortex creep, vortex lattice melting, a decrease in $T_c$ with an increase in an applied magnetic field, and the suppression of superconductivity at magnetic fields above the upper critical field $H_{c2}$;
4. Critical current density $j_c$, above which the resistance-less state disappears;
5. Superconducting energy gap indicating the collective nature of the charge carriers;
6. Pronounced isotope effect demonstrating the involvement of the crystal lattice in phonon-mediated superconductivity.

Studies of high-temperature superconductivity continue both theoretically, to better understand its high $T_c$ and the prospects for room-temperature superconductivity, and experimentally, with many new superconductors being discovered[6]. The initially discovered superconductors $H_3S$ and $LaH_{10}$ have been characterized most thoroughly, with key superconducting parameters such as $T_c$, Ginzburg-Landau parameter, energy gap, coherence length $\xi$, London penetration depth $\lambda_L$, upper critical field $H_{c2}$, lower critical field $H_{c1}$, thermodynamic critical field $H_c$, and critical current density $j_c$ estimated8.

Four-probe electrical transport measurements have been well developed for the use in diamond anvil cells and serve as a primary tool for detecting and studying superconductivity at high pressures. This method is routinely used in various laboratories [8,12-18] and references herein. Another independent technique for proving and investigating superconductivity is magnetic susceptibility measurements[5], which have recently been significantly improved for megabar high-pressure conditions[8]. Currently, three techniques are used for magnetic measurements at high pressures: a double modulation *AC* technique using a system of coils[19,20], SQUID magnetometry[5,21], and nitrogen-vacancy (NV) color centers implanted in diamond anvils as quantum sensors[22]. Additionally, the screening of applied magnetic fields by the $H_3S$ superconductor has been explored using the forward nuclear resonant scattering technique, employing the $^{119}Sn$ Mössbauer isotope as a sensor[23].

The double modulation *AC* approach measures an anomaly in magnetic response and provides a value of $T_c$ and a qualitative indication of the superconducting transition. This method is expected to be more widely used despite its relative complexity. This technique has detected the magnetic response arising from the superconducting nature of highly compressed hydrides such as $H_3S$[19] and $LaH_{10}$[20]. Another technique, static *DC* measurements using SQUID magnetometry, yields the absolute values of magnetic susceptibility. The development of miniature nonmagnetic diamond anvil cells[1,5,21] and the trapped flux method[24] have significantly expanded the applicability of SQUID magnetometry.

The NV centers technique has also been successfully adapted to the megabar pressure range, demonstrating high sensitivity and submicron spatial resolution. These features enable the detection of magnetic response from much smaller samples, for instance, few-micrometers-size $CeH_9$ phase[22]. Ongoing development of the NV center technique aims to increase the range of applied magnetic fields used for measurements.

Next, we turn our attention to the magnetic flux penetration field which was one of the measured quantities in our paper[1]. Currently, only *DC* measurements using SQUID magnetometry allow for the study of magnetic field penetration into a sample over a wide range of applied magnetic fields, enabling the estimation of $H_p$ (magnetic flux penetration field) values. $H_p$ is connected with the lower critical field, $H_{c1}$, via the demagnetization factor of the sample. The onset of applied magnetic field penetration into a superconductor is determined from the deviation of the virgin $m(H)$ magnetization curve from the linear response, which corresponds to the Meissner state.

To implement these measurements, a unique tiny DAC with a diameter of 8.8 mm was designed to perfectly fit the bore of commercial SQUIDs. The DAC body is made of materials with very low-magnetic susceptibility, yet it still significantly contributes to the recorded total signal. Nevertheless, the signal from the sample is clearly visible against the DAC background (see Figure 1a in [1]). This is remarkable because the mass of the DAC is approximately $10^8$ times larger than that of the sample.

The screening of low applied magnetic fields, indicating the Meissner state, and the presence of hysteretic loops upon sweeping high applied magnetic fields, evidencing the presence of persistent superconducting currents, were clearly observed in $H_3S$ samples in the pioneering work[5]. However, no initial $m(H)$ data, or so-called virgin curves, were measured, and $H_{c1}$ and $\lambda_L$ were only roughly estimated[5].

Additionally, $m(H)$ magnetization data of $H_3S$ at $P = 140$ GPa, including the virgin curve, were measured at a single temperature $T = 100$ K, using a magnetic property measurement system (MPMS) from Quantum Design, as reported in Ref.[21] At that time, we performed conventional cleaning of the DAC body, diamond anvils, seats, and gaskets to remove possible contamination stemming from the manufacture of these parts using instruments made of steel and tungsten carbide (with impurities of cobalt). However, the conventional cleaning did not completely eliminate the strong paramagnetic contribution of the DAC assembly, which significantly affected the measured total signal. It should also be noted that this DAC had additional elements (electrical leads, wiring, and solder) for electrical transport measurements, which also contributed to the background. Therefore, to minimize the paramagnetic perturbation, we performed these measurements using a standard background subtraction procedure of the MPMS software for the SQUID magnetometer, where the background was collected above $T_c$ at $T = 210$ K (see the inset in Figure 1). In our subsequent works[1,24], we succeeded in further development of post-machining cleaning procedures for the high-pressure DAC assembly (see below), and the background subtraction procedure was not applied for either $m(T)$ or $m(H)$ magnetic susceptibility measurements. More recently, the penetration of magnetic flux into $H_3S$ and $LaH_{10}$ superconductors was systematically studied using the almost background-independent trapped flux method over a wide temperature range[24].

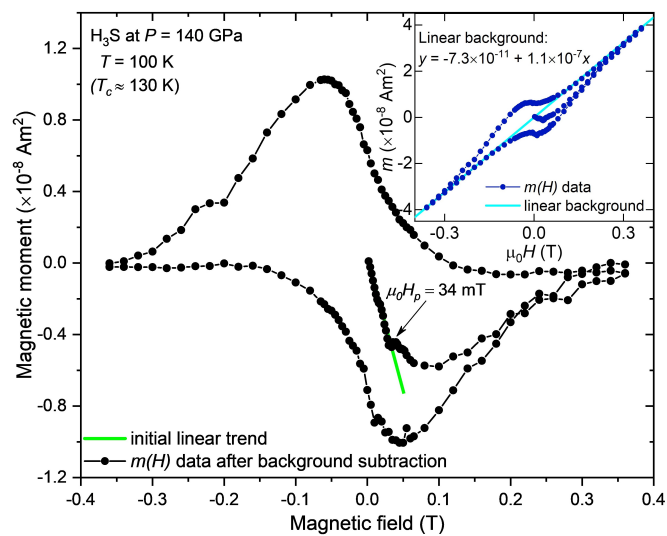

**Figure 1. Magnetization $m(H)$ measurements in $H_3S$ sample pressurized at $P = 140$ GPa ($T_c \approx 130$ K) from Ref.[21], demonstrating typical superconducting behavior.** Raw $m(H)$ data, including the background response from the DAC, are shown as blue circles in the inset. In this case, the background is paramagnetic, leading to a positive background slope in the raw data. After the subtraction of the background, approximated by a linear function (sky blue line in the inset), the $m(H)$ data exhibits the typical shape for a type-II superconductor. The virgin $m(H)$ curve shows a clear deviation from the initial linear trend (green line) at approximately 34 mT.

Our work[1] provided a significant step in understanding the behavior of $H_3S$ and $LaH_{10}$ samples in applied magnetic fields, demonstrating not only the qualitative features of superconducting phase in these compounds, such as screening and expulsion of the applied magnetic field and the presence of hysteretic loops upon sweeping applied magnetic fields at temperatures below $T_c$, but also the estimation of several superconducting parameters such as the penetration field $H_p$ and London penetration depth $\lambda_L$.

In our work[1] we made every effort to avoid all possible sources of contamination in our samples, which could lead to a paramagnetic background slope in magnetization $m(H)$ measurements. To achieve this, all elements of the high-pressure cell assembly were thoroughly etched in an ultrasonic cleaner with acids. All parts of the DACs and rhenium gaskets were etched in 3 M hydrochloric acid for 30 minutes, and the diamonds were etched in a mixture of concentrated nitric and hydrochloric acids in a 1:3 molar ratio for 90 minutes. As a result, the $m(H)$ data of DACs with the $Im$-$3m$-$H_3S$ and $Fm$-$3m$-$LaH_{10}$ samples were not distorted by paramagnetic slope from impurities, and showed the diamagnetic slope naturally expected from the magnetic susceptibility values of the materials used for the high pressure assembly (see Figure 2).

At a number of temperatures, we measured $m(H)$ data by sweeping applied magnetic fields in the range of -1 to 1 Tesla to obtain hysteretic loops at temperatures above and below $T_c$ of $H_3S$ and $LaH_{10}$. Several such data are shown in Figure 2. Since each $m(H)$ loop measurement took several days, at some temperatures in our study we looked at only the virgin curve which is the part of relevance to obtaining $H_p$.

We estimated $\mu_0 H_{c1}$ values of ~0.82 T and ~0.55 T, and $\lambda_L(0)$ of ~20 nm and ~30 nm in $H_3S$ and $LaH_{10}$, respectively, after considering the estimated geometry of the synthesized samples. The small values of $\lambda_L$ indicate a high superfluid density in both hydrides. These compounds have the values of the Ginzburg-Landau parameter $\kappa$ of ~12–20 and belong to the group of moderate-$\kappa$ type-II superconductors (e.g., $MgB_2$ and A15 alloys), rather than being hard superconductors, as might be intuitively expected from their high $T_c$s. Our later estimated value of $\lambda_L(10\ K) = 37$ nm in $H_3S$[24], does not alter this conclusion.

However, Hirsch, in his comment[2], expressed concerns about an inconsistency in the averaging of the data in our work[1]. In the end of his comment, Hirsch escalated his concerns to absurd scepticism about the entire field of high-$T_c$ conventional superconductivity.

A key point is how to determine the onset of the deviation of the virgin $m(H)$ part of the data from the linear trend in a case of noisy data that we had in minor instances. The source of the noise is the S700X SQUID magnetometer, which we used in all our experiments[1], it utilizes two current sources: the first provides magnetic fields up to 30 mT with low noise, and the second provides higher fields but with higher noise, especially in the field range of 35-150 mT. This design is standard from the manufacturer of the S700X SQUID magnetometer, and we cannot modify it. As a result, we had no problem with the signal-to-noise ratio and the estimation of $H_p$ values when the onset point of deviation from the linear trend lies within the range of magnetic fields < 30 mT. Fortunately, this holds true for all measured data for the $LaH_{10}$ sample (and for the majority of well-known superconductors) and partly for the $H_3S$ sample at $T = 180$ K and 160 K.

The deviation points for the $H_3S$ sample measured at lower temperatures shift to the range of noisy data. The physical origin for this shift is a large value of the lower critical field, $H_{c1}$, in $H_3S$ (or, in other words, short $\lambda_L$).

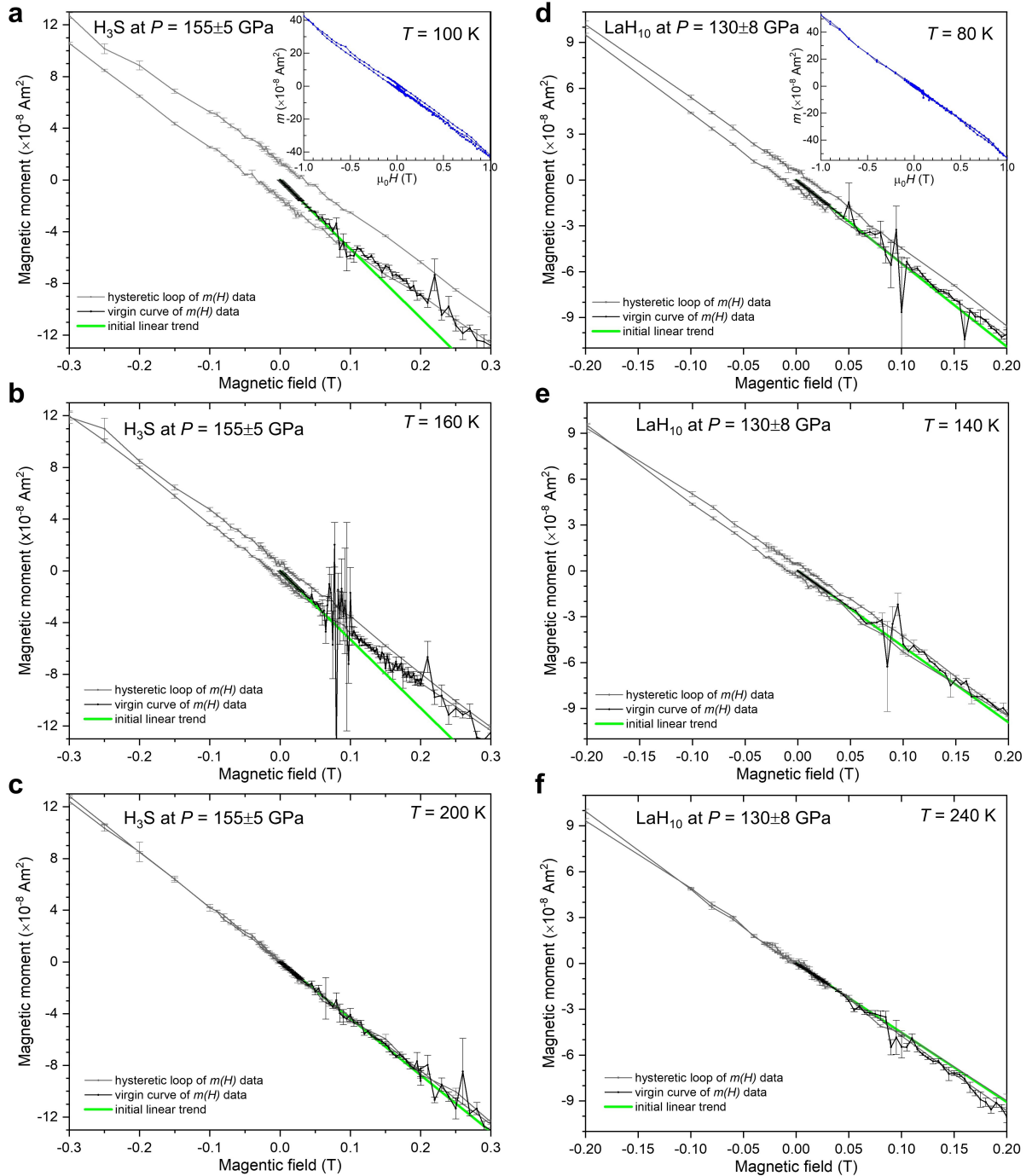

**Figure 2. Magnetization *m(H)* measurements in hydrogen-rich superconductors *Im-3m*-H₃S and *Fm-3m*-LaH₁₀.** *m(H)* data upon sweeping applied magnetic fields in the range of -1 to 1 T, including virgin curves, demonstrate: (**a**, **b**) the presence of characteristic superconducting hysteresis at $T$ = 100 K and 160 K (below $T_c$); and (**c**) the absence of the hysteresis at $T$ = 200 K (above $T_c$) in the *Im-3m*-H₃S sample pressurized at $P$ = 155±5 GPa. *m(H)* data upon sweeping applied magnetic fields in the range of -1 to 1 T, including virgin curves, demonstrate: (**d**, **e**) the presence of characteristic superconducting hysteresis at $T$ = 80 K and 140 K (below $T_c$); and (**f**) the absence of the hysteresis at $T$ = 240 K (above $T_c$) in the *Fm-3m*-LaH₁₀ sample pressurized at $P$ = 130±8 GPa. Black and grey circles correspond to the virgin and hysteretic portion of the *m(H)* data. Green lines represent the initial linear behaviour of the virgin *m(H)* data and serve as a guide to demonstrate the deviation of the virgin curve from the linear trend. The measured *m(H)* data showed a diamagnetic slope of the background of the rigorously etched DAC assembly. Insets demonstrate the whole measured range of *m(H)* datasets for the H₃S sample at $T$ = 100 K and the LaH₁₀ sample at $T$ = 80 K.

Even when the deviation is visible to the eye in the original scattered data, smoothing or averaging is required for an accurate and objective determination of $H_p$ values. Therefore, we first smoothed the original data in the noisy range to combine it with the low-noise data for the systematic determination of the deviation from linear behavior. This is not a trivial problem, which we solved in several steps of smoothing/averaging. In retrospect, we accept that the procedure we used is indeed open to criticism. Therefore, we report an alternative method for processing the original measured data to estimate $H_p$ without resorting to smoothing.

Since no standard approach to detecting $H_p$ is found in the literature, the deviation from the initial linear trend is typically determined by eye. Another source of uncertainty in the determination of $H_p$ in type-II superconductors, particularly those exhibiting strong pinning, arises from the absence of a dramatic change in the magnetic moment in virgin *m(H)* magnetization curves when the applied magnetic field exceeds $H_p$. This task is even more challenging for hydrogen-rich compounds, as their useful superconducting response is close to the sensitivity limit of the SQUID magnetometer and is further perturbed by the background signal of the diamond anvil cell (DAC).

We recently developed a generalized method for systematically determining the deviation of a magnetization curve from linear behavior. An identical approach is widely used for the determination of the critical current values in superconductors[3], and it has been successfully extended to extract $H_p$ from *DC* magnetization data[4] in various superconductors. This method is also attractive for us because datasets with variable noise levels can be easily analyzed.

The main idea of this method is to fit the measured $m(H_{appl})$ by the power-law function of the applied magnetic field $H_{appl}$. The power-law function includes a linear dependent term, which is attributed to the linear Meissner diamagnetic response. The deviation from the linear dependence is described by $m_c \times \left(\frac{H_{appl}}{H_p}\right)^n$ term, where $m_c$ is the threshold criterion to define the $H_p$. Thus, $m_c$ plays a similar role to the electric field criterion of $E_c = 1$ µV/cm used to determine the transport critical current density[3] from the *E–J* curves. The details of these methods can be found in Refs.[3,4] A distinct advantage of this method is that it provides a systematic criterion for detecting $H_p$ values based on phenomenologically established shape of the *m(H)* dependence in the vicinity of $H_p$, whereas previously employed methods rely on a threshold value that varies widely from report to report and are therefore much less consistent. The fitting function is described by Equation (1):

$$m(H_{appl}) = m_0 + k \times H_{appl} + m_c \times \left(\frac{H_{appl}}{H_p}\right)^n \qquad (1)$$

where $m_0$ is an instrumental offset, $k$ is a linear term (also known as the Meissner slope), $m_c$ is the threshold criterion, and $n$ is the power-law exponent.

It should be noted that a power-law equation is not applicable for fitting the entire *m(H)* magnetization curve and must be used solely to find the onset point where the measured magnetic moment starts to deviate from the initial linear trend due to the penetration of magnetic flux into the superconductor. For this reason, one should define $m_c$ and the range of $H_{appl}$ for reasonable fitting. Identical conditions exist in the determination of the critical current density from the *E–J* curves using the power-law approach.

For *m(H)* datasets where $H_p$ lies in the low-noise region, we set the $m_c$ criterion as approximately three times the value of the uncertainty of the measured magnetic moment, above which the deviation from the linear trend becomes reliable. We limited the range of *m(H)* data for the fitting to the value of $H_{appl}$ at which the difference between the measured $m(H_{appl})$ and the Meissner line (linear term) reached approximately ten times the value of $m_c$.

We show several such fits with fitting parameters in Figure 3 and Supplementary Figure 1. For better representation of the onset of $H_p$, we subtracted the linear background $y = a + bx$ ($a$ and $b$ parameters given for each plot). We also reanalyzed the value of $H_p$ for the $H_3S$ data at $P \approx 140$ GPa ($T_c \approx 140$ K) published in Ref.[21] The $H_p$ values refined by the new approach are consistent with those estimated in the original works[1,21].

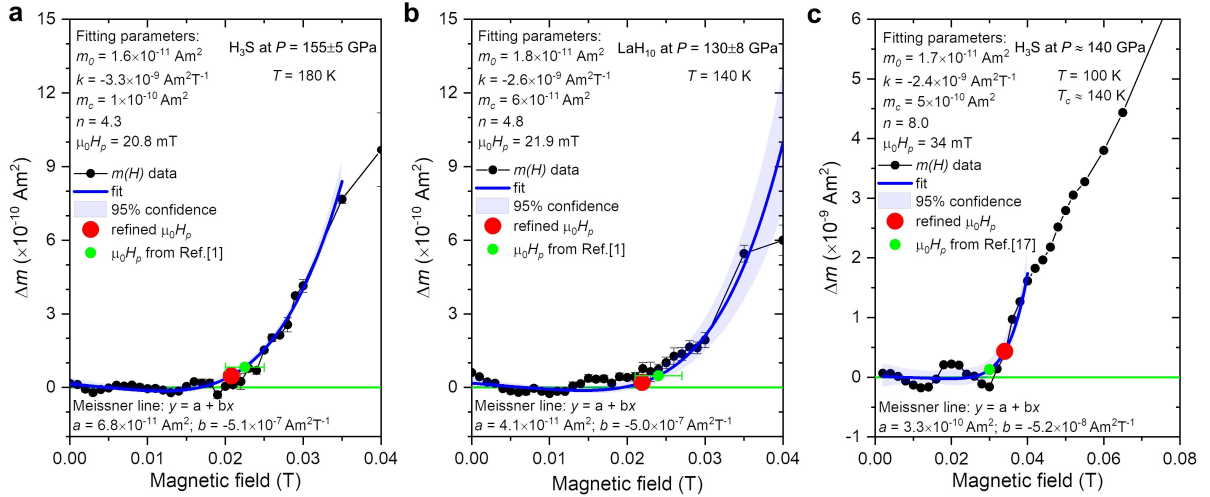

**Figure 3. The extraction of $H_p$ in the low-noise region of applied magnetic fields ($\leq 30$ mT).** (a) A portion of the $m(H)$ virgin curve for the *Im-3m*-$H_3S$ sample at $P = 155\pm5$ GPa, measured at $T = 180$ K from Ref.[1]; (b) a portion of the $m(H)$ virgin curve for the *Fm-3m*-$LaH_{10}$ sample at $P = 130\pm8$ GPa, measured at $T = 140$ K from Ref.[1]; (c) a portion of the $m(H)$ virgin curve for the $H_3S$ sample at $P \approx 140$ GPa ($T_c \approx 140$ K), measured at $T = 100$ K from Ref.[21] We subtracted the linear function of $y = a + bx$ (with a and b parameters shown below each curve) from the original $m(H)$ data for better presentation of the initial linear trend. This line represents the Meissner line and is defined as linear fit of the data measured at low applied magnetic fields below $H_p$. Black circles represent the measured $m(H)$ data; red and green circles represent $H_p$ values estimated in the present work and in Refs.[1,21], respectively. The green line depicts the Meissner line ($y = 0$), and blue curves represent the fits with 95% confidence bands.

For $m(H)$ datasets of the *Im-3m*-$H_3S$ sample at $P = 155\pm5$ GPa measured at $T \leq 140$ K, where $H_p$ values lie above 30 mT (in the noisy region), we set the $m_c$ criterion to approximately the averaged value of uncertainties of the $m(H_{appl})$ datasets used for fitting, including both low- and high-noise regions, but not larger than 10% of the maximum diamagnetic response $|\Delta m_{max}|$ from the superconducting $H_3S$ sample. Thus, for the dataset measured at $T = 140$ K, $m_c = 8\times10^{-10}$ Am$^2$ and for the datasets measured at $T \leq 120$ K, $m_c = 1\times10^{-9}$ Am$^2$. The data range for fitting was limited to magnetic fields at which the difference between the measured $m(H_{appl})$ and the Meissner line (linear term) reached approximately ten times the value of $m_c$, as it was done for the estimation of $H_p$ in the low-noise region. We show these fits with fitting parameters in Figure 4 and Supplementary Figure 2.

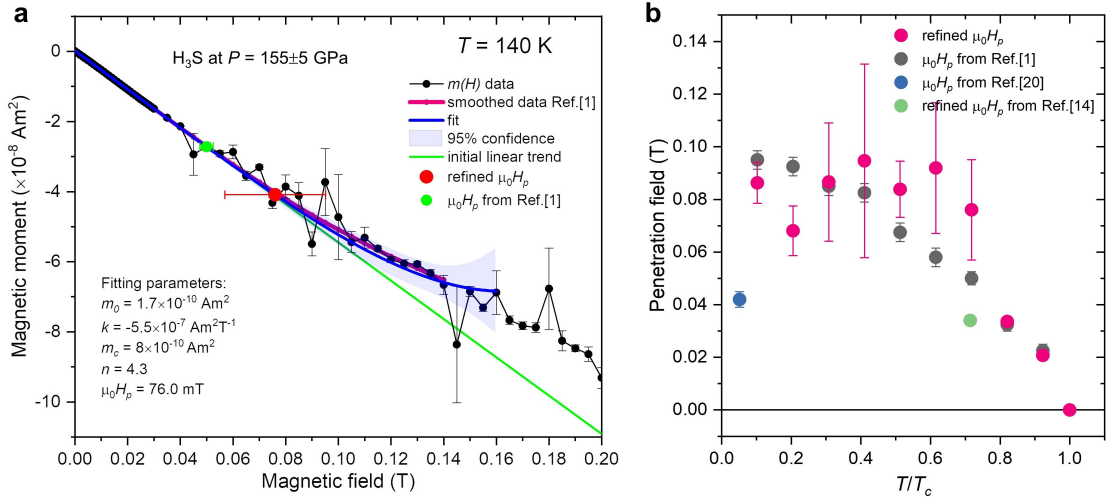

**Figure 4. The extraction of $H_p$ in the high-noise region of applied magnetic fields (> 30 mT) for the *Im-3m*-$H_3S$ sample at $P$ = 155±5 GPa.** (**a**) Fit of the portion of the $m(H)$ virgin curve for the *Im-3m*-$H_3S$ sample at $P$ = 155±5 GPa, measured at $T$ = 140 K by Equation (1). Fitting parameters are shown on the panel. The smoothed data from Ref.[1] are shown as magenta curve. The initial linear trend was defined as a linear fit of the measured data in the applied magnetic fields region of 0-20 mT. (**b**) Summary plot of the estimated $H_p$ values for $H_3S$ samples measured in different experiments: traditional magnetic susceptibility measurements (grey circles from Ref.[1] and green circle from Ref.[21]) and trapped magnetic flux method (blue circle, Ref.[24]).

In traditional magnetization measurements, the total magnetic signal from the pressurized sample unavoidably includes the response of the DAC body. The diamagnetic response of the DAC continuously grows with applied magnetic fields, while the useful signal from the superconducting sample weakens soon after $H_{appl}$ exceeds $H_p$. This leads to a decrease in the signal-to-background ratio at high applied magnetic fields, but does not affect the estimation of $H_p$ values.

Another important parameter is the background response of DACs to applied magnetic fields. Despite our efforts to manufacture miniature DACs with a background response to applied magnetic fields as low as possible, in practice, the background of the entire high-pressure assembly is not perfectly linear at low applied magnetics fields and across a wide temperature range. Since we cannot significantly increase the size of superconducting samples (the larger diamond anvil culets do not sustain these pressures) to improve the useful signal-to-background ratio, the subtle nonlinearity of the DAC background can manifest in the $m(H)$ measurements. However, this feature does not significantly affect the detection of superconductivity in our samples. Several cases for the background dependence of the DAC used in our work[1] at T = 20 K, 180 K, and 200 K are shown in Supplementary Figure 3. When the sample is in the normal state at 200 K, the virgin $m(H)$ curve shows a subtle downturn at approximately 10 mT. The same effect is observed for the data measured at lower temperatures, where $\mu_0 H_p$ values exceed 30 mT, i.e., when the nonlinear DAC background interferes with the Meissner state of the superconducting sample (linear magnetic field dependence of the $m(H_{appl})$). At temperatures where $\mu_0 H_p$ is below 30 mT, the effect of the penetration of the magnetic flux into the superconducting sample surpasses the nonlinearity in the DAC background, allowing for estimation of $H_p$.

We also checked the variation of $H_p$ values upon changing the range of the $m(H)$ dataset used for fitting. By extending the $H_{appl}$ range to two times ($\mu_0 H_{appl}$ = 300 mT) and three times ($\mu_0 H_{appl}$ = 440 mT) for $H_3S$ sample at $T$ = 100 K, the exponent of the fitting function reduces from $n$ = 5 to $n$ = 2.2, resulting in less accurate fitting of the experimental data. Despite this, the extracted $\mu_0 H_p$ values do not change significantly: from 84±10 mT to 63±9 mT, a difference of approximately 25%. The value of $\mu_0 H_p$ reported in Ref.[1] is 68±4 mT.

Applying the method proposed for the extraction of $H_p$ from *DC* magnetization measurements in Ref.[4] to our datasets for the *Im-3m*-$H_3S$ and *Fm-3m*-$LaH_{10}$ samples reported in Ref.[1], we obtained the refined values summarized in Figure 5. In particular, the refined $\mu_0H_p(0\,K) = 108\pm7$ mT for the *Im-3m*-$H_3S$ sample at $P = 155\pm5$ GPa and $35\pm3$ mT for the *Fm-3m*-$LaH_{10}$ sample at $P = 130\pm8$ GPa versus $97\pm2$ mT and $39\pm2$ mT for the datasets reported in Ref.[1]

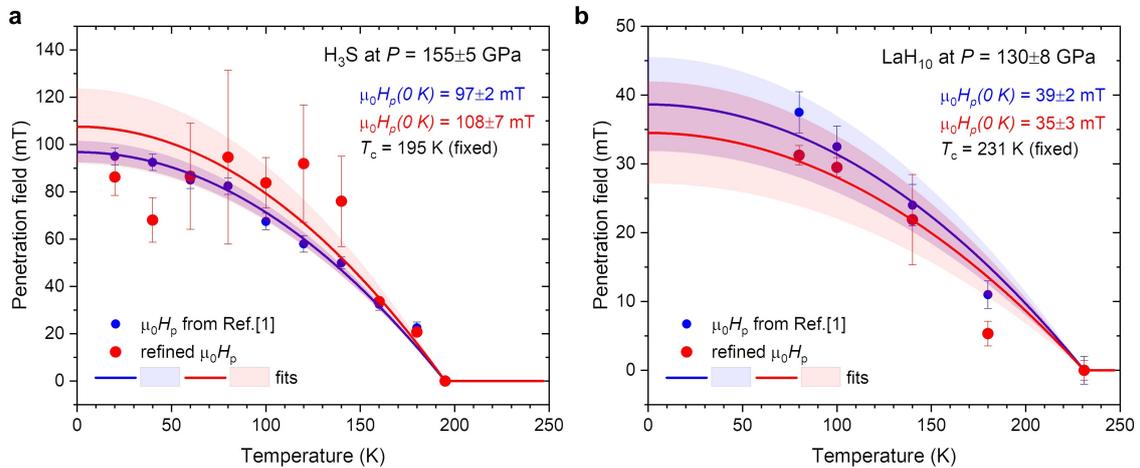

**Figure 5. Temperature dependence of estimated $H_p$ for the *Im-3m*-$H_3S$ and *Fm-3m*-$LaH_{10}$ samples. (a)** Refined $H_p$ values for the *Im-3m*-$H_3S$ sample at $P = 155\pm5$ GPa; and **(b)** refined $H_p$ values for the *Fm-3m*-$LaH_{10}$ sample at $P = 130\pm8$ GPa. $H_p$ values reported in Ref.[1] are shown grey circles. Fit of both datasets of estimated $H_p(T)$ values and their extrapolation to $H_p(0\,K)$ was done using equation $H_p(T) = H_p(0\,K) \times \left(1 - \left(\frac{T}{T_c}\right)^2\right)$. Blue and red circles represent the data form Ref.[1] and the refined data in the present work, respectively. Light blue and light red areas show 95% confidence bands of the fits.

It is noteworthy that estimating $H_{c1}(T)$ values necessitates knowing the demagnetizing factor, $N$. We derived the demagnetization correction, $(1-N)^{-1}$, from the estimated geometry of the synthesized samples, approximately 8.5 and 13.5 for $H_3S$ and $LaH_{10}$, respectively. These values assume that the superconducting phases are uniform and shaped as ideal thin solid disks. However, the real sample shape and continuity of the superconducting phase can differ from this ideal case, with the worst scenarios being the sample consisting of separated small crystallites, which could reduce the demagnetization correction to the value of 1.5, or the superconducting regions being thinner than the idealised disks, in which case it could be substantially larger than our estimates. The systematic errors on our estimates of $H_{c1}(T)$ are therefore substantial, but we believe that the estimate is worth attempting, because $\kappa \sim \sqrt{H_{c2}/H_{c1}}$ so a valuable estimate of $\kappa$ can be made even in the presence of large uncertainty in $H_{c1}$, and it is important to know approximately how strongly type-II the hydride superconductors are. The uncertainty in the real value of the demagnetizing factor is not specific to high-pressure measurements but also exists in ambient pressure studies, particularly for polycrystalline samples (e.g., see Figure 5 in Ref.[25]).

In summary, it is crucial to re-emphasize that we have significantly advanced experimental techniques for studying the properties of novel hydrogen-rich high-temperature superconductors at high pressures using SQUID magnetometry[1,5,21,23,24]. In particular, by utilizing traditional magnetic susceptibility[1,5,21,23] and new trapped flux[24] methods, we have demonstrated the presence of all 7 hallmarks of superconductivity in these compounds:

1) Screening and expulsion of the applied magnetic field by the superconducting phase below its $T_c$;

2) The appearance of hysteretic loops by sweeping the applied magnetic field at temperatures below $T_c$, related to persistent superconducting currents in superconductors and their growth with decreasing $T$;

3) The existence of an initial linear trend of *m(H)* in the virgin curve of magnetization data, related to the Meissner state of superconductors;

4) Trapping of magnetic flux by superconductors after switching off the applied magnetic field;

5) Different behaviours in the trapping of magnetic flux under zero-field-cooled and field-cooled protocols arising from the influence of the Meissner state of superconductors;

6) The presence of horizontal temperature-independent plateaus in $m_{trap}(T)$ measurements upon reverse cooling superconductors at temperatures below their $T_c$s;

7) Extremely slow creep of the trapped magnetic flux.

In addition, it should be mentioned that the reproducibility and consistency of our results between different magnetic measurements are notable. Both $H_3S$ and $LaH_{10}$ samples were also measured by trapped flux method[24], and the results obtained by two different techniques agree well with each other. Furthermore, the trapped flux method also provides the estimate of $\mu_0H_p$ (10 K) = 42±3 mT in $H_3S$ at $P$ = 155±5 GPa (diamond anvils in the DAC with $LaH_{10}$ sample were destroyed during measurements), which we consider more accurate since the background of the DAC did not perturb the response of the superconducting sample. The estimations of $H_p$ done by these two methods are consistent. It should be mentioned that in the literature, even ambient pressure superconductors demonstrate much larger dispersion in $H_p$.

In the present manuscript, we refined $H_p(T)$ values for the *Im-3m*-$H_3S$ and *Fm-3m*-$LaH_{10}$ samples by using an alternative method for determining the deviation from linear behaviour in applied magnetic fields in a more consistent manner. The refined $H_p(T)$ values are commensurate with the previously reported values in Ref.[1], and fitting curves of the $H_p(T)$ datasets to the Ginzburg-Landau critical field equation lie within the 95% confidence bands.

In addressing the broader suspicion raised about the validity of the whole hydride field of research by Hirsch and co-workers, we refer not just to our own work but to the multiple independent confirmations of hydride superconductivity by other groups, with the significant papers included in the citation list of this article.

## Acknowledgements

Authors are grateful to A. Mackenzie for helpful discussions. M.I.E. is thankful to the Max Planck community for valuable support and U. Pöschl for the encouragement. The National High Magnetic Field Laboratory is supported by the National Science Foundation through NSF/DMR-2128556, the State of Florida, and the U.S. Department of Energy. S.L.B. was supported by the U.S. Department of Energy, Office of Basic Energy Science, Division of Materials Sciences and Engineering under Contract No. DE-AC02-07CH11358. E.F.T. thanks financial support provided by the Ministry of Science and Higher Education of Russia within the theme "Pressure" No. 122021000032-5 and under Ural Federal University Program of Development within the Priority-2030 Program.

## Data Availability

The data supporting the findings of this study are openly available at the Open Science Framework (https://osf.io/7wqxb/) with DOI 10.17605/OSF.IO/7WQXB.

**Supplementary Materials**

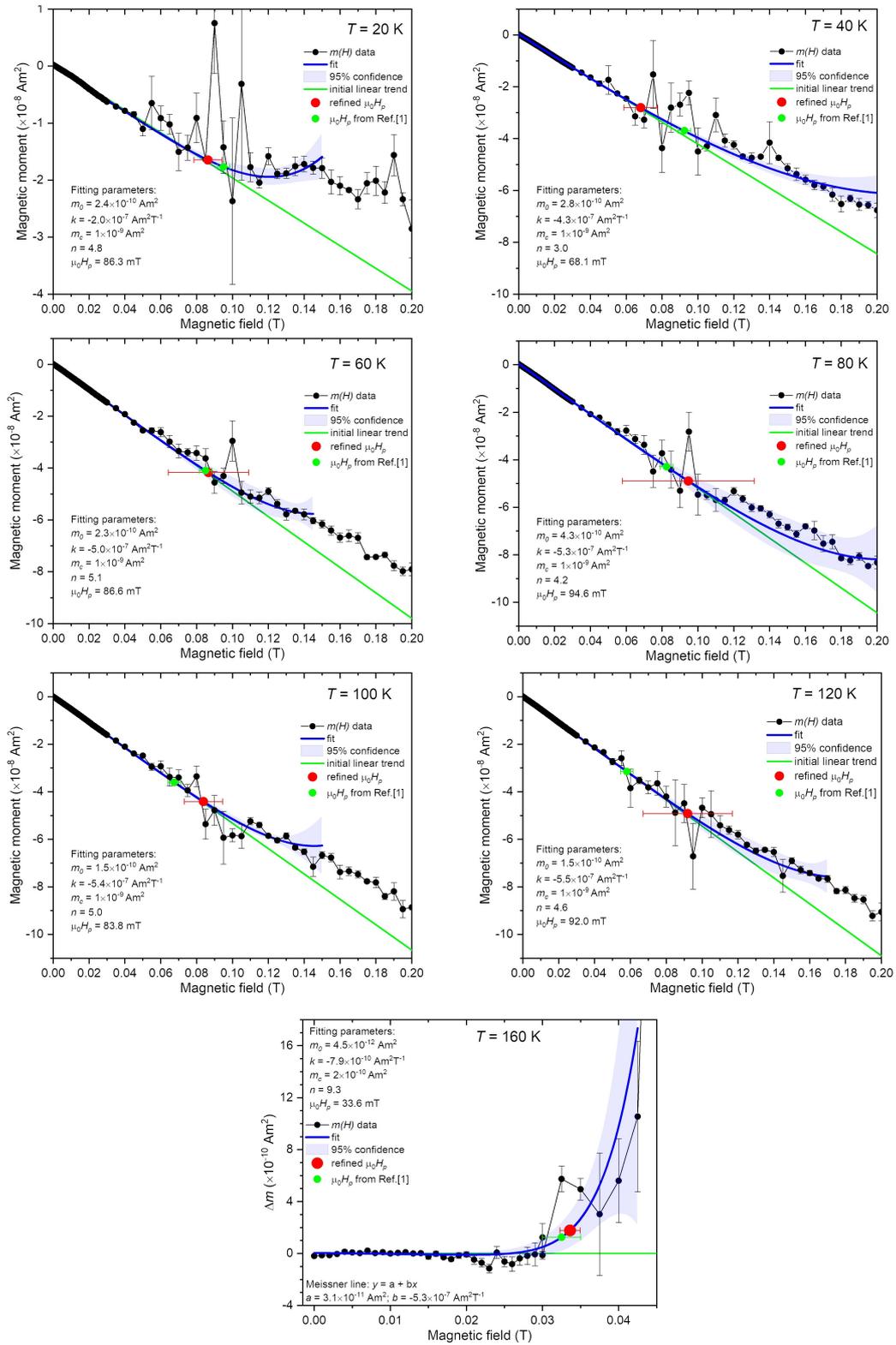

**Supplementary Figure 1. The extraction of $H_p$ from $m(H)$ magnetization data of the *Im-3m*-$H_3S$ sample at $P$ = 155±5 GPa.** Black circles represent the measured $m(H)$ data; red and green circles represent $H_p$ values estimated in the present work and in Ref.[1], respectively. The green line depicts the initial Meissner linear trend of the virgin curve in the Meissner state. This Meissner line is defined as linear fit of the data measured at low applied magnetic fields (≤ 30 mT) and serves as a guide for the eye. Blue curves represent fits with 95% confidence bands.

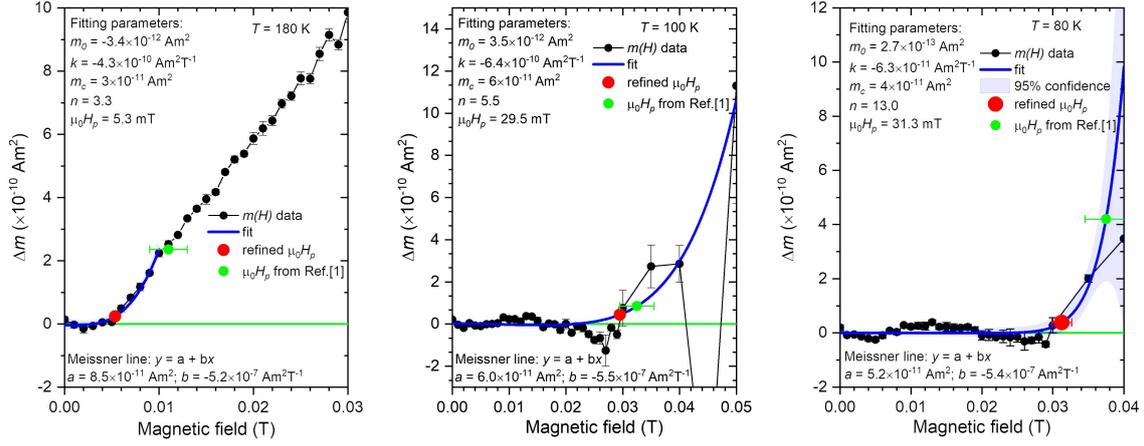

**Supplementary Figure 2. The extraction of $H_p$ from $m(H)$ magnetization data of the *Fm-3m*-LaH$_{10}$ sample at $P = 130 \pm 8$ GPa in the low-noise region of applied magnetic fields ($\leq 30$ mT).** We subtracted the linear function $y = a + bx$ (with a and b parameters shown below each curve) from the original $m(H)$ data for better presentation of the initial linear trend of the virgin curve. Black circles represent the measured $m(H)$ data; red and green circles represent $H_p$ values estimated in the present work and in Ref.[1], respectively. The green line depicts the Meissner line ($y = 0$), and blue curves represent the fits with 95% confidence bands.

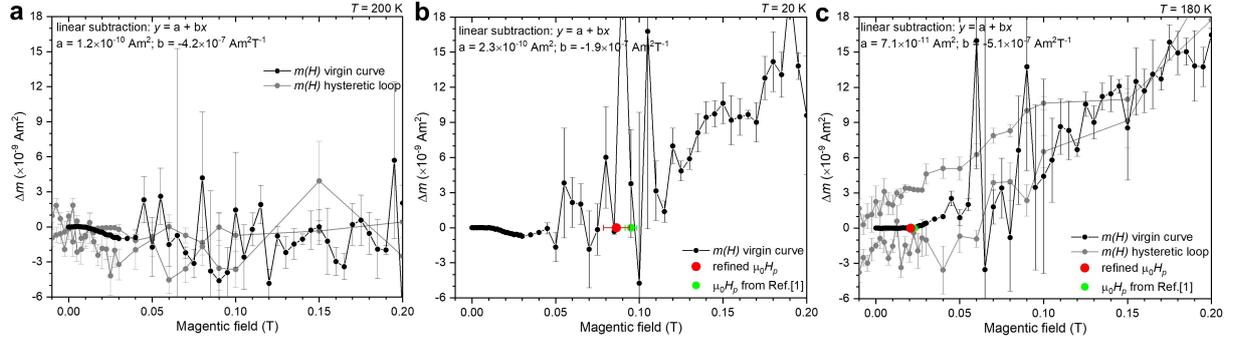

**Supplementary Figure 3. Nonlinearity of the background of the diamond anvil cell with the *Im-3m*-H$_3$S sample at $P = 155 \pm 5$ GPa in the region of applied magnetic fields below 30 mT.** (**a**) $m(H)$ data measured at $T = 200$ K (above $T_c$ of the superconducting H$_3$S sample) characterize the DAC background, showing a subtle downturn at approximately 10 mT. Importantly, this effect has the opposite sign compared to the penetration of magnetic flux into the superconducting sample at temperatures below $T_c$. (**b**) $m(H)$ data measured at $T = 20$ K demonstrate the same downturn from the DAC background, preceding the entrance of magnetic flux into the sample. This indicates that the $\mu_0 H_p$ value lies beyond 30 mT. (**c**) $m(H)$ data measured at $T = 180$ K indicate that the onset of magnetic flux penetration into the sample occurs at low applied magnetic fields (below 30 mT). In this case, the DAC background does not significantly affect the determination of the $H_p$ value. To better demonstrate the nonlinear behaviour of the DAC at low magnetic fields, we subtracted the linear background from the measured $m(H)$ data. The coefficients of the function $y = a + bx$ are obtained by linear fit to the very initial portion of the virgin $m(H)$ dataset ($0 < \mu_0 H_{appl} < 10$ mT) and are provided in each panel. Black circles represent the virgin portion of the $m(H)$ data; grey circles correspond to the hysteretic portion of the $m(H)$ data; green and red circles correspond to the refined values of $H_p$ (where applicable).

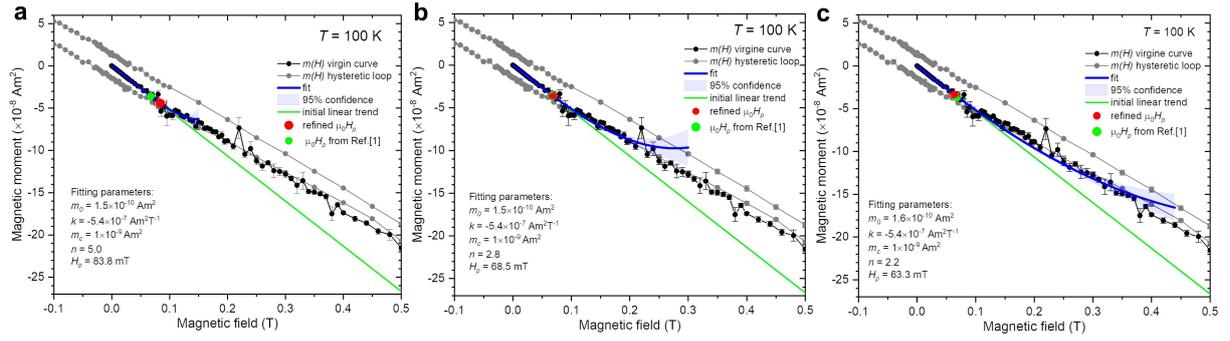

**Supplementary Figure 4. Influence of the range of the *m(H)* dataset of the *Im-3m*-H$_3$S sample (*P* = 155±5 GPa) measured at *T* = 100 K on *H$_p$* values upon fitting (when μ$_0$*H$_p$* lies in the region of applied magnetic fields above 30 mT).** (**a**) The fitting of the virgin curve of the *m(H)* data is limited by μ$_0$*H$_{appl}$* = 150 mT, at which the difference between the measured *m(H$_{appl}$)* and the Meissner line reaches approximately ten times the value of *m$_c$*. The refined μ$_0$*H$_p$* = 84±10 mT. (**b**) The fitting of the virgin *m(H)* data is extended to double the range, limited by μ$_0$*H$_{appl}$* = 300 mT. In this case, the refined μ$_0$*H$_p$* = 68±6 mT. (**c**) The results of the fitting when the selected *m(H)* dataset is limited by μ$_0$*H$_{appl}$* = 440 mT. The refined μ$_0$*H$_p$* = 63±9 mT. Black circles represent the measured *m(H)* data of the virgin magnetization curve; grey circles represent the hysteretic portion of the *m(H)* data; red and green circles represent *H$_p$* values estimated in the present work and in Ref.[1], respectively. The green line refers to the initial linear trend (Meissner line), and blue curves represent the fits with 95% confidence bands.